\documentclass[12pt,a4paper]{article}
\usepackage{amssymb}
\usepackage{amsfonts}
\usepackage[english]{babel}
\usepackage{graphicx}
\usepackage{appendix}
\newcommand{\ba}{\begin{array}}
\newcommand{\ea}{\end{array}}
\newcommand{\pa}{\partial}

\newcommand{\no}{\nonumber}

\newcommand{\be}{\begin{equation}}
\newcommand{\ee}{\end{equation}}
\newcommand{\bea}{\begin{eqnarray}}
\newcommand{\eea}{\end{eqnarray}}
\newcommand{\beaa}{\begin{eqnarray*}}
\newcommand{\eeaa}{\end{eqnarray*}}

\begin{document}

\title
{Soliton Interaction In the Modified Kadomtsev-Petviashvili-(II) Equation }
\author{
 Jen-Hsu Chang \\Department of Computer Science and Information Engineering, \\
 National Defense University, \\
 Tauyuan City, Taiwan, 33551 }

\date{}

\maketitle
\begin{abstract}

We study soliton interaction in the Modified Kadomtsev-Petviashvili-(II) equation (MKP-(II)) using the totally non-negative Grassmannian. One constructs the multi-kink soliton of MKP equation using the $\tau$-function and the Binet-Cauchy formula, and then investigates the interaction between kink solitons and line solitons. Especially, Y-type kink-soliton resonance, O-type kink soliton and  P-type kink soliton of X-shape are investigated. Their amplitudes of interaction are computed after choosing appropriate phases.

Keywords: $\tau$-function, Grassmannian, Resonance, Kink Soliton

\end{abstract}

\newpage

\section{Introduction}
Recently, the soliton interaction  in integrable models has attracted much attention, especially the resonant theory in the KP-(II) theory \cite{bc, ko,ko1,ko2, ko3, ko5, ko6} (references therein) and Novikov-Veselov (NV) equation \cite{jh1, jh2}.   The key point of soliton interaction in the KP-(II) equation is the $\tau$-function structure, i.e., the Wronskian formula. Using the Bitnet-Cauchy formula, one can express the $\tau$-function as a linear combination of exponential functions, whose coefficients have to satisfy the Plucker relations.   To get the non-singular solutions, one  leads to the totally non-negative Grassmannian. Then we can classify the resonant structures of KP-(II) equation using the Grassmannian. Due to the success in KP-(II) equation, one can investigate the soliton interaction of the Modified KP-(II) (MKP-(II)) equation.  The $\tau$-function structure of MKP-(II) is the same as the KP-(II) equation; however, the solution of MKP-(II) equation is associated with  the quotient of $\tau$-functions,  i.e., there is a Miura transformation between the solutions of KP-(II) equation and MKP-(II) equation (see below). To get the non-singular solutions of MKP-(II) equation, the parameters are non-negative. In particular, one has kink-soliton solution and then can investigate their resonant structure. \\
\indent  The MKP-(II) equation  is defined by \cite{dk, kn, kd, jm}
 \bea
-4u_{t}+u_{xxx}-6u^2u_{x}+6u_{x}\pa_x^{-1}u_y+3\pa_x^{-1}u_{yy}=0. \label{mkp}
 \eea
The equation (\ref{mkp}) was introduced in \cite{kn} within the framework of gauge-invariant description of the KP equation. In \cite{jm}, it appeared as the first member of modified KP hierarchy using the $\tau$-function theory. In \cite{kd}, the inverse-scattering-transformation  method  is used to get the exact solution for MKP equations, including rational solutions (lumps), line solitons and breathers. The MKP-(II) equation (\ref{mkp}) may be relevant to the description of water waves in a situation when one has to take cubic non-linearity into account. Also, it has been obtained by solving the associated coupled Maxwell and Landau-Lifshitz equations in two dimensions using a reductive perturbation method during the study on the propagation of electromagnetic wave (EMW) in an isotropic charge-free infinite ferromagnetic thin film \cite{vd}. In \cite{vd} it has pointed out that the magnetization of the medium is excited in the form of solitons and also the magnetic field component of the propagating EMW is modulated in the form of solitons. The MKP-(II) equation has also been derived in the study of the propagation of ion-acoustic waves in a plasma with 
non-isothermal electrons \cite{xz}. This model can also describe the evolution of various solitary waves in the multi-temperature electrons plasmas, in which there exists a collision-less multi-component plasma conceiving cold ions and two temperature electrons having different Maxwellian distributions rendered in the form of two Boltzmann relations \cite{ds} . \\
\indent Letting 
\be u(x,y,t)= \pa_x \ln (F(x,y,t)/G(x,y,t)), \label{tr}
\ee
we have the Hirota bi-linear equation \cite{hi, jm}
\bea
&& (D_y-D_x^2)F \circ G =0  \label{h1} \\
&& (-4D_t +D_x^3+3 D_x D_y)F \circ G =0  \label{h2}, 
\eea
where the bi-linear operators $D_x^m$ and $D_y^n$ are defined by 
\[D_x^m D_y^n F \circ G = (\pa_x-\pa_{x^{'}})^{m} (\pa_y-\pa_{y^{'}})^{n} F(x,y)G(x^{'},y^{'}).\]
To construct the solutions of these Hirota equations (\ref{h1}) and(\ref{h2}), one defines the determinant 
\be
\tau_N^{(n)}= det 
\left[\ba{cccc} f_1^{(n)}  & f_1^{(n+1)} & \cdots   &  f_1^{(n+N-1)}    \\
 f_2^{(n)} &  f_2^{(n+1)} & \cdots   &   f_2^{(n+N-1)} \\
\vdots  & \vdots & \vdots & \vdots    \\
 f_N^{(n)}& f_N^{(n+1)} & \cdots  &   f_N^{(n+N-1)}   \ea \right],    \ee
where the elements in the above determinant are defined by ($ i=1,2,3 \cdots, N $ )
\be  
\frac{\pa f_i}{\pa x_m}  = \frac{\pa^m f_i}{ \pa x^m} , \quad x_1=x, \quad x_2=y, \quad x_3=t,  \label{lin} 
\ee
and $f_i^{(n)}$ means the n-th order derivative with respect to $x$, $n=0,1,2,3, \cdots$. Also, we can write $\tau_N^{(n)}$ as a Wronskian, i.e., 
\[ \tau_N^{(n)}=Wr( f_1^{(n)}, f_2^{(n)}, f_3^{(n)}, \cdots, f_N^{(n)} ). \]
It is shown that \cite{hi}
\be
F=\tau_N^{(1)}, \quad G=\tau_N^{(0)}, \quad or \quad  u(x,y,t)= \pa_x \ln \frac{\tau_N^{(1)}}{\tau_N^{(0)}} , \label{h3}
\ee 
will be  solutions of (\ref{h1}) and (\ref{h2}) for $N=1,2, \cdots $. \\
\indent We remark that after the Miura transformation \cite{kd}, using the Hirota equation (\ref{h1}), we have 
\bea 
v= - \pa_x^{-1}u_y-u_x-u^2=2\pa_{xx} \ln \tau_N^{(0)}, \label{mi}
\eea 
and then one can obtain the Hirota equation by (\ref{h2}) 
\[(-4D_t D_x +D_x^4+3  D_y^3) \tau_N^{(0)} \circ  \tau_N^{(0)}=0, \]
or the KP-(II) equation 
\bea 
-4v_t+v_{xxx}+6vv_x+\pa_x^{-1}3v_{yy}=0. \label{kp}
\eea
\indent Next, we construct the resonant solutions of MKP-(II) equation 
using the totally non-negative Grassmannian in KP-(II) theory \cite{  ko3, ko6}. Here one considers a finite dimensional solution
\beaa 
f_i (x,y,t)  &=& \sum_{j=1}^M a_{ij} E_j (x,y,t), \quad i=1,2, \cdots N < M, \\
E_j (x,y,t) &= & e^{\theta_j}, \quad  \theta_j= k_j x+k_j^2 y+ k_j^3 t + \xi_j, \quad j=1,2, \cdots M 
\eeaa 
where $k_j $ and $\xi_j$ are real parameters. For simplicity, we take $\xi_j=0$ in this article. Each $E_j(x,y,t)$ satisfies the equations (\ref{lin}). Then each resonant solution of MKP-(II) equation can be parametrized  by a full rank matrix 
\[ A= \left[\ba{cccc} a_{11}  & a_{12}  & \cdots   &  a_{1M}   \\
 a_{21} &  a_{22} & \cdots   &   a_{2M} \\
\vdots  & \vdots & \vdots & \vdots    \\
 a_{N1} & a_{N2}  & \cdots  &   a_{NM}    \ea \right] \in M_{N \times M} (\textbf{R}). \]  
Using the Binet-Cauchy formula, the $\tau$-function $\tau_N^{(0)}$ can be written as 
\bea 
\tau_A &=& \tau_N^{(0)} = Wr (f_1, f_2, \cdots, f_N)=det 
\left[\ba{cccc} f_1 & f_1^{'} & \cdots   &  f_1^{(N-1)}    \\
 f_2 &  f_2^{'} & \cdots   &   f_2^{(N-1)} \\
\vdots  & \vdots & \vdots & \vdots    \\
 f_N & f_N^{'} & \cdots  &   f_N^{(N-1)}   \ea \right]  \no \\
&=& det \left [\left(\ba{cccc} a_{11}  & a_{12}  & \cdots   &  a_{1M}   \\
 a_{21} &  a_{22} & \cdots   &   a_{2M} \\
\vdots  & \vdots & \vdots & \vdots    \\
 a_{N1} & a_{N2}  & \cdots  &   a_{NM}    \ea \right) \left(\ba{cccc} E_1  & k_1 E_1 & \cdots   &  k_1^{N-1}E_1   \\
 E_2 & k_2 E_2 & \cdots   &   k_2^{N-1} E_2 \\
\vdots  & \vdots & \vdots & \vdots    \\
 E_M & k_M E_M  & \cdots  &    k_M^{N-1} E_M   \ea \right)  \right] \no \\
&=& \sum_J \Delta_J (A) E_J (x,y,t), \label{ta}
\eea 
where $\Delta_J (A)$ is the  $ N \times N $ minor for the columns  with the index set $J=\{ j_1, j_2, j_3, \cdots, j_N\} $, and $E_J$ is the Wronskian 
\be  E_J= Wr (E_{j_1},E_{j_2}, E_{j_3}, \cdots, E_{j_N} )= \prod_{m< l} (k_{j_l}-k_{j_m}) E_{j_1} E_{j_2}E_{j_3} \cdots E_{j_N}. \label{va} \ee
We notice that the coefficients $ \Delta_J (A)$ of $\tau_A$  have to satisfy the Plucker relations. \\
\indent Similarly, 
\bea 
\tau_A^{(1)} &=& \tau_N^{(1)} = Wr (f_1^{'}, f_2^{'}, \cdots, f_N^{'}) =det 
\left[\ba{cccc} f_1^{'} & f_1^{''} & \cdots   &  f_1^{(N)}    \\
 f_2^{'} &  f_2^{''} & \cdots   &   f_2^{(N)} \\
\vdots  & \vdots & \vdots & \vdots    \\
 f_N^{'} & f_N^{''} & \cdots  &   f_N^{(N)}   \ea \right]  \no \\
&=& det \left [\left(\ba{cccc} k_1 a_{11}  & k_2 a_{12}  & \cdots   & k_M a_{1M}   \\
k_1  a_{21} &  k_2 a_{22} & \cdots   &   k_M a_{2M} \\
\vdots  & \vdots & \vdots & \vdots    \\
 k_1 a_{N1} & k_2 a_{N2}  & \cdots  &  k_M  a_{NM}    \ea \right) \left(\ba{cccc} E_1  & k_1 E_1 & \cdots   &  k_1^{N-1}E_1   \\
 E_2 & k_2 E_2 & \cdots   &   k_2^{N-1} E_2 \\
\vdots  & \vdots & \vdots & \vdots    \\
 E_M & k_M E_M  & \cdots  &    k_M^{N-1} E_M   \ea \right)  \right] \no \\
&=& \sum_J \Delta_J (A) k_{j_1}  k_{j_2}  k_{j_3} \cdots    k_{j_N} E_J (x,y,t).  \label{taa}
\eea 
To obtain non-singular solutions of MKP-(II), from (\ref{h3}),  (\ref{ta}) and  (\ref{taa}), it can be seen that  $ \Delta_J (A) \geq 0$ for all $J$, i.e.,  A is an element of  totally non-negative Grassmannian $Gr(N,M)$, and we assume the ordering in the $k$-parameters, 
\be  0 \leq k_1 < k_2 < k_3 < \cdots < k_M. \label{or} \ee
We remark here that the ordering 
\be  k_1 < k_2 < k_3 < \cdots < k_M. \label{or1} \ee
can obtain singular solutions of MKP-(II); however, it can obtain non-singular solutions of KP-(II) after the Miura transformation (\ref{mi}). The solutions of MKP-(II) under the condition (\ref{or}) are called Type II solutions, and they are pure 2+1 dimensional ones. On the other hand,  the solutions of MKP-(II) under the condition (\ref{or1}) are called Type I solutions, and they admit 1+1 dimensional reduction \cite{kd}. For example, there is no solution of modified KdV (MKdV) equation obtained from the condition (\ref{or}). \\
\indent This paper is organized as follows: in section 2, we construct basic resonant solutions and then terrace-type solutions can be found.  In section 3, we investigate the X-shape solitons, i.e., O-type and P-type solitons.  The maximum amplitudes of the intersection of X-shape solitons are computed; moreover, the amplitudes of interaction between line soliton and kink soliton are found. In section 4,  we conclude the paper with several remarks.

\section{Basic Resonant Solutions}
In this section, one constructs basic resonant solutions. We study the resonant interaction between line soliton and kink soliton, and find out terrace-type solutions. The resonant interaction inside the kink fronts is studied.  In addition, the asymptotic line solitons are described as $ y \to \pm \infty $. \\
\indent Let's consider one line soliton. For N=1, one takes 
\[f_1=E_1+ E_2=2e^{(\theta_1+ \theta_2)/2} \cosh\frac{\theta_2-\theta_1}{2}. \]
Also, 
\[f_{1x}=k_1 E_1+ k_2 E_2= 2e^{(\Theta_1+ \Theta_2)/2} \cosh\frac{\Theta_2-\Theta_1}{2},  \]
where \[   \Theta_j=\theta_j+ \ln k_j.  \]
Then one can get the line soliton: 
\be u=\pa_x \ln \frac{f_{1x}}{f_1}= \frac{k_2-k_1}{2} (\tanh \frac{\theta_2-\theta_1+\ln \frac{k_2}{k_1} }{2}-\tanh \frac{\theta_2-\theta_1}{2}) \geq 0. \label{pr} \ee 
A simple calculation shows that when $ \theta_2-\theta_1=-\frac{1}{2} \ln \frac{k_2}{k_1}$, $u$ has maximal value $ (\sqrt{k_2}-\sqrt{k_1})^2. $ Similar to the case KP-(II) \cite{ko1}, it can be seen that 
the $[1,2]$-line soliton solution (\ref{pr}) has the wave vector 
\be \vec {K}_{[i,j]}=(k_j-k_i, k_j^2-k_i^2),  \quad i=1, \quad j=2,  \label{tr} \ee 
and can be measured in the counterclockwise sense from the $y$-axis, i.e., 
\be \tan \Phi_{[i,j]}=\frac{k_j^2-k_i^2}{k_j-k_i}=k_i+k_j, \quad i=1, \quad j=2 ; \label{po} \ee
moreover, its velocity is given by 
\be \vec {V}_{[i,j]}= \frac{k_i^2+k_ik_j+k_j^2}{1+(k_i+k_j)^2} (1, k_i+k_j), \quad i=1, \quad j=2,  \label{ve} \ee
and the frequency is given by
\be \Omega_{i,j}= k_j^3-k_i^3=(k_j-k_i)(k_i^2+k_ik_j+k_j^2), \quad i=1, \quad j=2.  \label{fe} \ee
From (\ref{ve}), we see that any soliton propagates in the positive $x$-direction. \\
\indent We notice that a  kink solution can be obtained by $k_1=0$. In this case, from (\ref{pr}), we have 
\be u=\frac{k_2}{2} (1-\tanh \frac{\theta_2}{2} ) \to  \left\{\ba {ll} k_2 , &  x   \to -\infty ,  \\   0 ,  &  x  \to \infty . \ea \right.  \label{hii} \ee
One defines that the kink front of (\ref{hii}) is $ \theta_2=0$. Then by (\ref{po}) and (\ref{ve}) its wave vector and velocity are 
\be \tan \Phi_{[0,j]}=k_j, \quad \vec {V}_{[0,j]}= \frac{k_j^2}{1+k_j^2} (1, k_j), \quad  \Omega_{0,j}= k_j^3,  \quad j=2, \label{k2} \ee
respectively. \\
\indent From the form of $\tau$-function (\ref{ta}), the $xy$-plane is partitioned into several regions depending on the dominant exponential $E_J$ in its own region. Each line soliton is obtained by the balance between adjacent regions and is localized only at the boundaries of the dominant regions. In \cite{bc}, it is proved that as $\vert y \vert \to \infty $ the unbounded line solitons  remain invariant for any fixed time in KP-(II) equation case. From (\ref{ta}) and (\ref{taa}), it can be seen locally 
\beaa 
\tau_A & \approx & E_{i, j_2,  j_3   \cdots, j_N}+ E_{j , j_2,  j_3   \cdots, j_N}  \\
\tau_A^{(1)} & \approx & k_i k_{j_2} k_{j_3} \cdots k_{j_N} E_{i, j_2,  j_3   \cdots, j_N} + k_j k_{j_2} k_{j_3} \cdots k_{j_N} E_{j, j_2,  j_3   \cdots, j_N}. 
\eeaa
A similar calculation as (\ref{pr}) can yield locally ( $ [i,j]$-soliton )
\be  u \approx  \frac{k_j-k_i}{2} (\tanh \frac{\Theta_j-\Theta_i+\ln \frac{k_j}{k_i} }{2}-\tanh \frac{\Theta_j-\Theta_i}{2}) \geq 0, \label{prg} \ee
where 
\[ \Theta_j=\theta_j+ \ln \vert \prod_{m=2}^N (k_j-k_{j_m})\vert , \quad \Theta_i=\theta_i+ \ln \vert \prod_{m=2}^N (k_i-k_{j_m})\vert . \]
Also, when $ \Theta_j-\Theta_i=-\frac{1}{2} \ln \frac{k_j}{k_i}$, this $ [i,j]$-soliton has maximal value $ (\sqrt{k_j}-\sqrt{k_i})^2$.  It has the wave vector (\ref{tr}), the velocity (\ref{ve}) and the frequency (\ref{fe}).  \\
\indent  A  multi-kink solilton can be obtained by $k_1=0$. In this case, each  $[1,j]$-line soliton in  (\ref{prg}) becomes kink front, i.e., 
\be  u \approx \frac{k_j}{2} (1-\tanh \frac{\Theta_j- \ln  \prod_{m=2}^N k_{1_m} }{2} ) \to  \left\{\ba {ll} k_j , &  x   \to -\infty ,  \\   0 ,  &  x  \to \infty, \ea \right.  \label{mk} \ee
and forms the boundary of the multi-kink solution. The front of the multi-kink solution of (\ref{mk}) is defined as
\[ \Theta_j- \ln  \prod_{m=2}^N k_{1_m} =0.  \]
Their wave vectors , the velocities  and the frequencies (\ref{fe}) are defined by (\ref{k2}).  \\
\indent Next, we consider basic resonant solitons, i.e., Y- type solutions. Firstly, for N=1, one takes
\be g_1=E_1+ E_2+ E_3,  \quad u=\pa_x \ln \frac{g_{1x}}{g_1}=\pa_x \ln \frac{k_1E_1+ k_2 E_2+k_3 E_3}{E_1+ E_2+ E_3} . \label{th} \ee
Similar to the KP-(II) equation \cite{ko1}, three line solitons can interact to form a trivalent vertex and satisfy the resonant conditions for wave number and frequency by (\ref{tr}) and (\ref{fe}) ($i < m < j $)
\be  \vec{K}_{[i,j]}= \vec{K}_{[i,m]}+ \vec{K}_{[m,j]}, \quad   {\Omega}_{[i,j]} ={\Omega}_{[i,m]} + {\Omega}_{[m,j]}, i=1, m=2, j=3.   \label{re} \ee
\indent For the kink soliton of (\ref{th}), we take $k_1=0$. Please see the figure 1.
\begin{figure}[h]
	\includegraphics[width=1.2\textwidth]{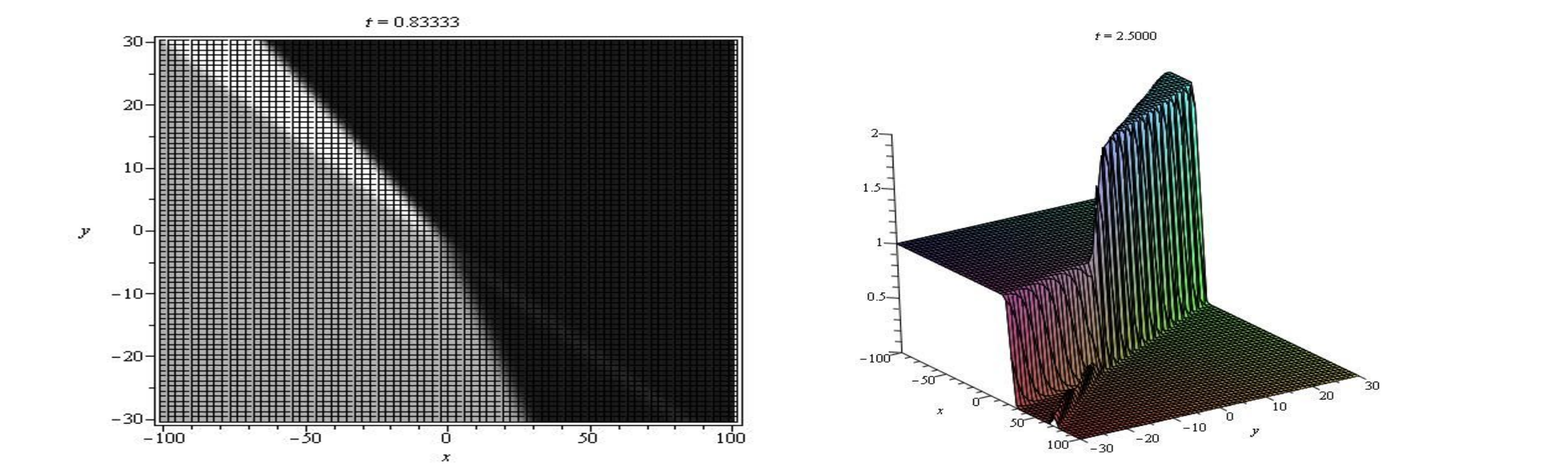}
	\caption{{ Y-type Kink (I)     ($k_1=0, k_2=1, k_3=2$  ) }}
\end{figure} One obtains two kink fronts (from left to right):  
\begin{itemize}
\item For $ y >> 0$: one has  $[2,3]$-front  and $[1,3]$-front. 
\item For $ y <<  0$:  one has one kink front $[1,2]$-front and the line soliton $[2,3]$-soliton. 
\end{itemize}
From (\ref{pr}), the kink bounded by $[2,3]$-front  and $[1,3]$-front has the height: $\frac{k_3}{2} (1-\tanh \frac{\theta_3}{2} ) $, and the kink bounded by $[2,3]$-front  and $[1,2]$-front has the height: $ \frac{k_2}{2} (1-\tanh \frac{\theta_2}{2} )$. The wave vectors and  velocities of  $[1,3]$-front and $[1,2]$-front are given by (\ref{k2}) for $j=3,2$. These three fronts satisfy the resonant conditions (\ref{re}). Also, we notice that  $[2,3]$ is both a front and line soliton, its wave vector and velocity being by (\ref{po}) and (\ref{ve}), $i=2, j=3$. We see that the line soliton $[2,3]$-soliton penetrates through the kink soliton and becomes the boundary of different height of kink solitons. It is different from the KP-(II) case. \\
\indent Secondly, one considers another basic Y-type soliton for $N=2$. We consider the matrix
\[ A_Y=\left[\ba{ccc} 1 & 0 & -b    \\ 0 & 1 & a   \ea
 \right]. \]
 where $a,b$ are positive number. The we know 
\[f_1=E_1-b E_3, \quad f_2=E_2+a E_3.  \]
 By the formula (\ref{ta}), the corresponding $\tau$-function is
 \[ \tau_{A_Y}= Wr(f_1, f_2)= (k_2-k_1) E_1E_2+a (k_3-k_1)E_1E_3+b (k_3-k_2)E_2E_3. \]
So 
\bea u &=& \pa_x \ln \frac{Wr(f_1', f_2')}{Wr(f_1, f_2)} \no \\
&=& \pa_x \ln \frac{k_1k_2 (k_2-k_1) E_1E_2+a k_1k_3(k_3-k_1)E_1E_3+b k_2k_3(k_3-k_2)E_2E_3}{(k_2-k_1) E_1E_2+a (k_3-k_1)E_1E_3+b (k_3-k_2)E_2E_3} . \label{ree} \eea
For the kink soliton of (\ref{ree}), we take $k_1=0$. Please see the figure 2. 
\begin{figure}[h]
	\includegraphics[width=1.2\textwidth]{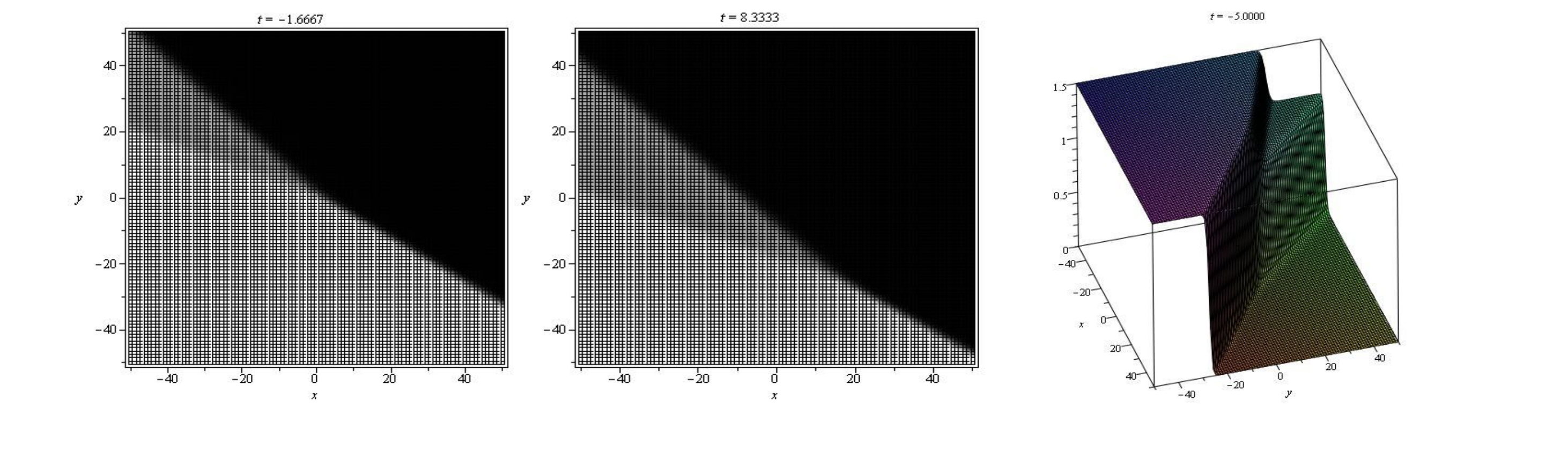}
	\caption{{ Y-type Kink (II)     ($k_1=0, k_2=1, k_3=2, a=10, b=40 $  ) }}
\end{figure}
One obtains two kink fronts 
\begin{itemize}
\item for $ y >> 0$: $[2,3]$-front  and $[1,2]$-front  (from left to right); 
\item for $ y <<  0$ : one has one kink front $[1,3]$-front. 
\end{itemize} 
Also, from (\ref{pr}), the kink bounded by $[2,3]$-front  and $[1,2]$-front has the height: $\frac{k_2}{2} (1-\tanh \frac{\theta_2}{2} ) $, and the kink bounded by $[2,3]$-front  and $[1,3]$-front has the height: $ \frac{k_3}{2} (1-\tanh \frac{\theta_3}{2} )$. The wave vectors and  velocities of  $[1,3]$-front and  $[1,2]$-front are given by (\ref{k2}) for $j=3,2$ but the wave vector and  velocity of  $[2,3]$-front is given by (\ref{po}) and (\ref{ve}). These three fronts also  satisfy the resonant conditions (\ref{re}). \\
\indent We remark here that in \cite {hz,jt, sc} the Y-type resonance of line solitons of MKP-(II) is investigated but their N-soliton solutions are different from (\ref{taa}), i.e., there is no non-negative Grassmannian structure or Plucker relations.

\section{X-type  Solitons}
In this section, one constructs O-type and P-type solitons using the totally non-negative Grassmannian. Then the amplitudes of intersection of line solitons are computed after choosing appropriate phases. Furthermore, the interaction between line soliton and kink soliton is described. \\

\subsection{O-type soliton}
The  Grassmannian of O-type has the form \cite{ko1}
\[ A_O=\left[\ba{cccc} 1 & a & 0 & 0  \\ 0 & 0 & 1 & b  \ea
 \right], \]
where  $a, b$ are positive numbers. 
Then the $\tau$-function  is
\[ \tau_O = (k_3-k_1) e^{\theta_1+ \theta_3}+ b (k_4-k_1)e^{\theta_1+ \theta_4}+ a (k_3-k_2)e^{\theta_2+ \theta_3}+ ab (k_4-k_2) e^{\theta_2+ \theta_4}. \]
And \[ \tau_O^{(1)}  = k_1 k_3 (k_3-k_1) e^{\theta_1+ \theta_3}+ b k_1 k_4 (k_4-k_1)e^{\theta_1+ \theta_4}+ a k_2 k_3(k_3-k_2)e^{\theta_2+ \theta_3} + ab k_2 k_4 (k_4-k_2) e^{\theta_2+ \theta_4}. \]

Near [1,2] soliton,  we have by (\ref{prg}) 
\be u \approx \frac{k_2-k_1}{2}(\tanh \frac{\Theta_2- \Theta_1+ \ln \frac{k_2}{k_1}}{2}-\tanh \frac{\Theta_2- \Theta_1}{2}), \label{uu} \ee
where 
\[ \Theta_2- \Theta_1=\theta_2- \theta_1+ \ln a +\ln (k_4-k_2)-\ln (k_4-k_1).\]
When $ \Theta_2- \Theta_1= -\frac{1}{2} \ln \frac{k_2}{k_1}$, we have the maximal amplitude. Also, the phase shift is :
\beaa
\theta_{[1,2]}^+ &=& \frac{1}{2} \ln \frac{k_2}{k_1}-\ln a+ \ln (k_4-k_1)-\ln (k_4-k_2) \\
\theta_{[1,2]}^- &=& \frac{1}{2} \ln \frac{k_2}{k_1}-\ln a+ \ln (k_3-k_1)-\ln (k_3-k_2). 
\eeaa
Then 
\[  \theta_{[1,2]}= \theta_{[1,2]}^+ - \theta_{[1,2]}^-= \ln \Delta_O, \]
where 
 \[\Delta_O= \frac{ (k_4-k_1)(k_3-k_2) }{ (k_4-k_2)( k_3-k_1) }=1- \frac{ (k_2-k_1)(k_4-k_3) }{ (k_4-k_2)( k_3-k_1) }.\]
 Notice that $ 0 \leq \Delta_O \leq 1.$  Then  $\theta_{[1,2]}=  \theta_{[3,4]} < 0 $. Each  $[i,j]$-soliton shifts  in $x$ with 
\[\Delta x_{i,j}= \frac{1}{k_j-k_i}\theta_{[i,j]} <0, \]
 which indicates an attractive force in the interaction \cite{ko}.\\
\indent Now, we can choose $a$ such that 
\be \theta_{[1,2]}^+ + \theta_{[1,2]}^-= \ln \frac{k_2}{k_1}+ \ln \frac{k_4-k_1}{k_4-k_2}-2 \ln a + \ln \frac{k_3-k_1}{k_3-k_2}=0. \label{ca} \ee
Then 
\[ a=\sqrt{\frac{k_2(k_4-k_1)(k_3-k_1)}{k_1(k_4-k_2)(k_3-k_2)}}. \]
Likewise, near [3,4] soliton, one yields 
\[ u \approx \frac{k_4-k_3}{2}(\tanh \frac{\Theta_4- \Theta_3+ \ln \frac{k_4}{k_3}}{2}-\tanh \frac{\Theta_4- \Theta_3}{2}), \]
where 
\[ \Theta_4- \Theta_3=\theta_4- \theta_3+ \ln b +\ln (k_4-k_1)-\ln (k_3-k_1).\]
When $  \Theta_4- \Theta_3= -\frac{1}{2} \ln \frac{k_4}{k_3}$, we have the maximal amplitude. Also, the phase shift is 
\beaa
\theta_{[3,4]}^+ &=& \frac{1}{2} \ln \frac{k_4}{k_3}-\ln b+ \ln (k_3-k_2)-\ln (k_4-k_2) \\
\theta_{[3,4]}^- &=& \frac{1}{2} \ln \frac{k_2}{k_1}-\ln b+ \ln (k_3-k_1)-\ln (k_4-k_1). 
\eeaa
Then 
\[ \theta_{[3,4]}= \theta_{[3,4]}^+ - \theta_{[3,4]}^-= \ln \Delta_O=\theta_{[1,2]}. \]
We can choose $b$ such that 
\[\theta_{[3,4]}^+ + \theta_{[3,4]}^-= \ln \frac{k_4}{k_3}+ \ln \frac{k_3-k_1}{k_4-k_1}-2 \ln b + \ln \frac{k_3-k_2}{k_4-k_2}=0. \]
Then 
\[ b=\sqrt{\frac{k_4(k_3-k_2)(k_3-k_1)}{k_3(k_4-k_2)(k_4-k_1)}}. \] 
For these  particular choices of $a$ and $b$, we have using (\ref{h3}), after a little algebra,  
\beaa
 \tau_O &\equiv & \sqrt{k_1k_3} e^{\theta_1+ \theta_3}+ \sqrt{k_1 k_4 \Delta_O} e^{\theta_1+ \theta_4}+ \sqrt{k_2 k_3 \Delta_O}e^{\theta_2+ \theta_3} + \sqrt{k_2 k_4}e^{\theta_2+ \theta_4} \\
&=&  e^{\hat \theta_1+ \hat \theta_3}+ \sqrt{\Delta_O} e^{\hat \theta_1+ \hat \theta_4}+ \sqrt{\Delta_O}e^{\hat \theta_2+ \hat \theta_3} + e^{\hat \theta_2+ \hat \theta_4} \\
&=& e^{\hat \theta_1+\hat \theta_2+ \hat \theta_3+ \hat \theta_4} [E_{12}^+ ( E_{34}^+ + \sqrt{\Delta_O}E_{34}^-) + E_{12}^- ( \sqrt{\Delta_O} E_{34}^+ + E_{34}^-)] \\
&\equiv &  \cosh  {\hat \theta_+}^O +\sqrt{\Delta_O} \cosh  {\hat \theta_-}^O, \\
\eeaa
where
\beaa
 {\hat \theta_{\pm}}^O   &=& \frac{1}{2} [  ({\hat \theta_2}- {\hat \theta_1}) \pm ({\hat \theta_4}-  {\hat \theta_3})]  \\
           E_{ij}^{ \pm} &=& e^{\frac{\pm (\hat \theta_i-\hat \theta_j)}{2} }, \quad 
					 \hat \theta_j = \theta_j +\frac{1}{2} \ln k_j,
\eeaa
and $ \equiv $ means it is equivalent by (\ref{h3}). 
Similarly, one has 
\[  \tau_O^{(1)} \equiv    \cosh  {\tilde \theta_+}^O +\sqrt{\Delta_O} \cosh  {\tilde \theta_-}^O, \] 
where 
 \[\quad  {\tilde \theta_{\pm}}^O = \frac{1}{2} [  ({\tilde \theta_2}- {\tilde \theta_1}) \pm ({\tilde \theta_4}-  {\tilde \theta_3})], \quad  \tilde \theta_j=\theta_j +\frac{3}{2} \ln k_j. \]
Then 
\[ u= \pa_x \ln \frac{ \tau_O^{(1)}}{ \tau_O }= \pa_x \ln \frac{ \cosh  {\tilde \theta_+}^O+\sqrt{\Delta_O} \cosh  {\tilde \theta_-}^O }{\cosh  {\hat \theta_+}^O+\sqrt{\Delta_O} \cosh {\hat \theta_-}^O}.\] 
\indent Next, we compute  the amplitude of the intersection part of $[1,2]$-soliton and $[3,4]$-soliton . It is determined  by the linear system
\bea
{ \hat \theta_2}- {\hat \theta_1} &=&  -\frac{1}{2} \ln \frac{k_1}{k_2} \no \\  
{ \hat \theta_4}- {\hat \theta_3} &=&  -\frac{1}{2} \ln \frac{k_4}{k_3}. \label{int} 
\eea
Then 
\beaa
{\hat \theta_+^O} &=& -\frac{1}{4} (\ln \frac{k_1}{k_2}+ \ln \frac{k_4}{k_3})  \\
{\hat \theta_-^O} &=& -\frac{1}{4} (\ln \frac{k_1}{k_2}- \ln \frac{k_4}{k_3}), 
\eeaa
and 
\beaa
{\tilde \theta_+^O} &=& \frac{1}{4} (\ln \frac{k_1}{k_2}+ \ln \frac{k_4}{k_3})  \\
{\tilde \theta_-^O} &=& \frac{1}{4} (\ln \frac{k_1}{k_2}- \ln \frac{k_4}{k_3}).  
\eeaa   
These imply 
\[  \tilde \theta_+^O=- \hat \theta_+^O , \quad \tilde \theta_-^O=- \hat \theta_-^O. \]
A direct calculation shows that at the intersection part 
\bea
u &=& (k_2-k_1)\frac{(\sqrt{k_2k_4}-\sqrt{k_1k_3})+ \sqrt{\Delta_O}(\sqrt{k_2k_3}-\sqrt{k_1k_4}) }{(\sqrt{k_2k_4}+\sqrt{k_1k_3})+ \sqrt{\Delta_O}(\sqrt{k_2k_3}+\sqrt{k_1k_4}) } \no \\
  &+& (k_4-k_3)\frac{(\sqrt{k_2k_4}-\sqrt{k_1k_3})- \sqrt{\Delta_O}(\sqrt{k_2k_3}-\sqrt{k_1k_4}) }{(\sqrt{k_2k_4}+\sqrt{k_1k_3})+ \sqrt{\Delta_O}(\sqrt{k_2k_3}+\sqrt{k_1k_4}) }. \label{am}
\eea 
We remark that in \cite{lc} the N-soliton solution of MKP-(II) is constructed using the Darboux transformation and the O-type soliton of two lines  is investigated; however, the authors didn't compute the amplitude of intersection of these two line solitons. 

Using the inequality, $  0<x=\sqrt{\Delta_O}<1$, $a,b,c,d \in R $, 
\[  \frac{a+b}{c+d} < \frac{ax+b}{cx+d} < \frac{b}{d}, \quad iff \quad (ad-bc) < 0, \]
it is not difficult to see that 
\[  A_{[1,2]} + A_{[3,4]} < u < \frac{(\sqrt{k_2k_4}-\sqrt{k_1k_3}) }{(\sqrt{k_2k_4}+\sqrt{k_1k_3})} (k_4-k_3+k_2-k_1). \]
The middle portion has the highest amplitude.  Also, 
\begin{itemize} 
\item when $\Delta_O=1$, we get $k_3=k_4$ or $k_2=k_1$. Then $u=A_{[1,2]}$ or $ A_{[3,4]} $, i.e., one-line soliton;  
\item when $\Delta_O=0$, we get $k_3=k_2$. Then $u=A_{[1,4]}=(\sqrt{k_4}-\sqrt{k_1})^2$, i.e., Y-type soliton; moreover, if $ A_{[1,2]}= A_{[3,4]}=A $, then we have $u=4A$.  It is similar to the KP-(II) case  . 
\item when $k_1=0$, from (\ref{uu}) and (\ref{ca}) we choose $a$ such that,  noticing that the kink front is  $\Theta_2- \Theta_1=0$,   
\[ \theta_{[1,2]}^+ + \theta_{[1,2]}^-= \ln \frac{k_4}{k_4-k_2}-2 \ln a + \ln \frac{k_3}{k_3-k_2}=0. \]
Then 
\beaa \tau_O &=&  \sqrt{k_3} e^{\theta_1+ \theta_3}+ \sqrt{\Delta_O} \sqrt{k_4} e^{\theta_1+ \theta_4}+ \sqrt{\Delta_O} \sqrt{k_3}e^{\theta_2+ \theta_3}\\
 &+&   \sqrt{k_4} e^{\theta_2+ \theta_4}. \eeaa
And $   \tau_O^{(1)} =   k_2 k_3  \sqrt{\Delta_O} \sqrt{k_3}e^{\theta_2+ \theta_3} +  k_2 k_4 \sqrt{k_4} e^{\theta_2+ \theta_4}. $ Then also from (\ref{am})
\be  u=k_2+ {\frac{(k_4-k_3)^2}{(\sqrt{k_4-k_2}+ \sqrt{k_3-k_2} )^2}}.   \label{ki} \ee
The last term is the interaction term between line soliton and kink soliton. Please see the  figure 3. 
\end{itemize}  
\begin{figure}[h]
		\includegraphics[width=0.9\textwidth]{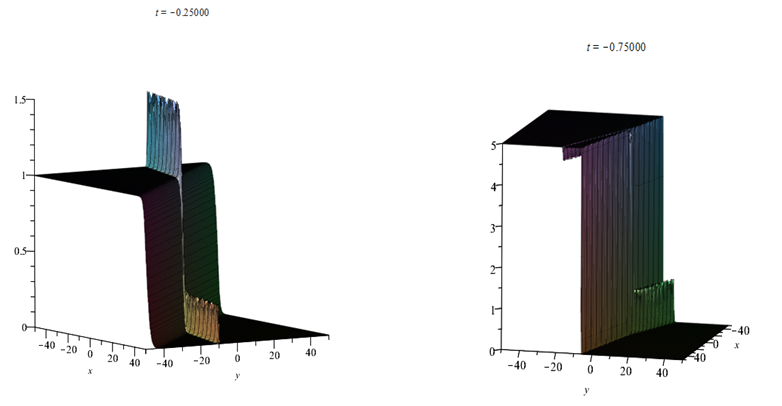}
	\caption{Left: O-type Kink Soliton ($ k_1=0, k_2=1, k_3=1.5, k_4=3 $  ); Right: P-type Kink Soliton($ k_1=0, k_2=0.3, k_3=2.5, k_4=5$  )}
\end{figure} 

\subsection{P-type Soliton}
The Grassmannian of  the P-type has the form \cite{ko1}
\[ A_P=\left[\ba{cccc} 1 & 0 & 0 & -b  \\ 0 & 1 & a & 0  \ea
 \right], \]
where  $a, b$ are positive numbers. 
Then the $\tau$-function  is
\[ \tau_P = (k_2-k_1) e^{\theta_1+ \theta_2}+ a (k_3-k_1)e^{\theta_1+ \theta_3}+ b(k_4-k_2)e^{\theta_2+ \theta_4}+ ab (k_4-k_3) e^{\theta_3+ \theta_4}. \] 
And \[ \tau_P^{(1)} = k_1 k_2 (k_2-k_1) e^{\theta_1+ \theta_2}+ a k_1 k_3 (k_3-k_1)e^{\theta_1+ \theta_3}+ b k_2 k_4(k_4-k_2)e^{\theta_2+ \theta_4} + ab k_3 k_4 (k_4-k_3) e^{\theta_3+ \theta_4}. \]
Near [1,4] soliton, one has by (\ref{prg})
\be u \approx \frac{k_4-k_1}{2}(\tanh \frac{\Theta_4- \Theta_1+ \ln \frac{k_4}{k_1}}{2}-\tanh \frac{\Theta_4- \Theta_1}{2}), \label{up} \ee
where 
\[ \Theta_4- \Theta_1=\theta_4- \theta_1+ \ln b +\ln (k_4-k_3)-\ln (k_3-k_1).\]
When $ \Theta_4- \Theta_1= -\frac{1}{2} \ln \frac{k_4}{k_1}$, we have the maximal amplitude. Also, the phase shift is :
\beaa
\theta_{[1,4]}^+ &=& \frac{1}{2} \ln \frac{k_4}{k_1}-\ln b+ \ln (k_3-k_1)-\ln (k_4-k_3) \\
\theta_{[1,4]}^- &=& \frac{1}{2} \ln \frac{k_4}{k_1}-\ln b+ \ln (k_2-k_1)-\ln (k_4-k_2). 
\eeaa
Then 
\[  \theta_{[1,4]}= \theta_{[1,4]}^+ - \theta_{[1,4]}^-=\ln (\Delta_P), \]
where $ \Delta_P= \frac{ (k_4-k_2)( k_3-k_1) }{(k_2-k_1)(k_4-k_3) }. $  Notice that $  \Delta_P \geq 1 .$ Then $\theta_{[1,4]}=  \theta_{[2,3]} > 0 $. Each  $[i,j]$-soliton shifts  in $x$ with 
\[ \Delta x_{i,j}= \frac{1}{k_j-k_i}\theta_{[i,j]} > 0, \]
 which indicates an repulsive force in the interaction \cite{ko}.   \\
\indent Now, we can choose $b$ such that 
\be \theta_{[1,4]}^+ + \theta_{[1,4]}^-= \ln \frac{k_4}{k_1}+ \ln \frac{k_3-k_1}{k_4-k_3}-2 \ln b + \ln \frac{k_2-k_1}{k_4-k_2}=0. \label{cp} \ee
Then 
\[ b=\sqrt{\frac{k_4(k_2-k_1)(k_3-k_1)}{k_1(k_4-k_3)(k_4-k_2)}}. \]
Similarly, near [2,3] soliton, one yields
\[ u \approx \frac{k_3-k_2}{2}(\tanh \frac{\Theta_3- \Theta_2+ \ln \frac{k_2}{k_3}}{2}-\tanh \frac{\Theta_3- \Theta_2}{2}), \]
where 
\[ \Theta_3- \Theta_2=\theta_3- \theta_2+ \ln a +\ln (k_4-k_3)-\ln (k_4-k_2).\]
When $  \Theta_3- \Theta_2= -\frac{1}{2} \ln \frac{k_3}{k_2}$, we have the maximal amplitude. Also, the phase shift is 
\beaa
\theta_{[2,3]}^+ &=& \frac{1}{2} \ln \frac{k_3}{k_2}-\ln a+ \ln (k_4-k_2)-\ln (k_4-k_3) \\
\theta_{[2,3]}^- &=& \frac{1}{2} \ln \frac{k_3}{k_2}-\ln a+ \ln (k_2-k_1)-\ln (k_3-k_1). 
\eeaa
Then 
\[ \theta_{[2,3]}= \theta_{[2,3]}^+ - \theta_{[2,3]}^-= \ln \Delta_P=\theta_{[1,4]}. \]
We can choose $a$ such that 
\[\theta_{[2,3]}^+ + \theta_{[2,3]}^-= \ln \frac{k_3}{k_2}+ \ln \frac{k_2-k_1}{k_3-k_1}-2 \ln a + \ln \frac{k_4-k_2}{k_4-k_3}=0. \]
Then 
\[ a=\sqrt{\frac{k_3(k_2-k_1)(k_4-k_2)}{k_2(k_3-k_1)(k_4-k_3)}}. \] 
Similar to the O-type soliton, for these  particular choices of $a$ and $b$ , we have, after a simple calculation, 
\beaa
 \tau_P &\equiv &  \cosh  {\hat \theta_+}^P+\sqrt{\Delta_P} \cosh  {\hat \theta_-}^P \\
 \tau_P^{(1)} &\equiv &   \cosh  {\tilde \theta_+}^P+ \sqrt{\Delta_P} \cosh  {\tilde \theta_-}^P,  
\eeaa
where
\beaa
 {\hat \theta_{\pm}}^P   &=& \frac{1}{2} [  ({\hat \theta_4}- {\hat \theta_1}) \pm ({\hat \theta_3}-  {\hat \theta_2})]  \\
 {\tilde \theta_{\pm}}^P   &=& \frac{1}{2} [  ({\tilde \theta_4}- {\tilde \theta_1}) \pm ({\tilde \theta_3}-  {\tilde \theta_2})],  
\eeaa
and $ \hat \theta_j=\theta_j +\frac{1}{2} \ln k_j, \quad \tilde \theta_j=\theta_j +\frac{3}{2} \ln k_j. $
Then 
\[ u= \pa_x \ln \frac{ \tau_P^{(1)}}{ \tau_P }=\pa_x \ln \frac{ \cosh  {\tilde \theta_+}^P+ \sqrt{\Delta_P} \cosh  {\tilde \theta_-}^P}{\cosh  {\hat \theta_+}^P+\sqrt{\Delta_P}  \cosh  {\hat \theta_-}^P}.\] 
\indent Next, we compute  the amplitude of the intersection part of $[1,4]$-soliton and $[2,3]$-soliton . It is determined  by the linear system
\bea
{ \hat \theta_3}- {\hat \theta_2} &=&  -\frac{1}{2} \ln \frac{k_3}{k_2} \no \\  
{ \hat \theta_4}- {\hat \theta_1} &=&  -\frac{1}{2} \ln \frac{k_4}{k_1}. \label{inp} 
\eea
Then 
\beaa
{\hat \theta_+^P} &=& -\frac{1}{4} (\ln \frac{k_3}{k_2}+ \ln \frac{k_4}{k_1})  \\
{\hat \theta_-^P} &=& -\frac{1}{4} (\ln \frac{k_3}{k_2}- \ln \frac{k_4}{k_1}), 
\eeaa
and 
\beaa
{\tilde \theta_+^P} &=& \frac{1}{4} (\ln \frac{k_3}{k_2}+ \ln \frac{k_4}{k_1})  \\
{\tilde \theta_-^P} &=& \frac{1}{4} (\ln \frac{k_3}{k_2}- \ln \frac{k_4}{k_1}).  
\eeaa   
These imply 
\[  \tilde \theta_+^P=- \hat \theta_+^P, \quad \tilde \theta_-^P=- \hat \theta_-^P. \]
A direct calculation shows that at the intersection part 
\bea
u &=& (k_4-k_1)\frac{(\sqrt{k_3k_4}-\sqrt{k_1k_2})+ \sqrt{\Delta_P}(\sqrt{k_4k_2}-\sqrt{k_1k_3}) }{(\sqrt{k_3k_4}+\sqrt{k_1k_2})+ \sqrt{\Delta_P}(\sqrt{k_2k_4}+\sqrt{k_1k_3}) } \no \\
  &+& (k_3-k_2)\frac{(\sqrt{k_3k_4}-\sqrt{k_1k_2})- \sqrt{\Delta_P}(\sqrt{k_4k_2}-\sqrt{k_1k_2}) }{(\sqrt{k_3k_4}+\sqrt{k_1k_2})+ \sqrt{\Delta_P}(\sqrt{k_2k_4}+\sqrt{k_1k_3})}. \label{amp}
\eea 
Using the inequality, $  0<x=\frac{1}{\sqrt{\Delta_P}} <1$, $a,b,c,d \in R $, 
\[  \frac{b}{d} < \frac{ax+b}{cx+d} < \frac{a+b}{c+d}, \quad iff \quad (ad-bc) > 0, \]
it is not difficult to see that 
\[ \frac{(\sqrt{k_2k_4}-\sqrt{k_1k_3}) }{(\sqrt{k_2k_4}+\sqrt{k_1k_3})} (k_4-k_3+k_2-k_1) < u <  A_{[1,4]} + A_{[2,3]}. \]
The middle portion has the lower amplitude  than the [1,4]-line soliton's one. It is very different from the O-type soliton. Also, 
\begin{itemize} 
\item when $\Delta_P=\infty$, we get $k_3=k_4$ or $k_2=k_1$. Then $u=A_{[1,2]}$ or $ A_{[3,4]} $, i.e., Y-type soliton;  
\item when $\Delta_P=1$, we get $k_3=k_2$. Then $u=A_{[1,4]}$, i.e., one-line soliton. 
\item when $k_1=0$, from (\ref{up}) and (\ref{cp}) we choose $b$ such that, noticing that the kink front is  $\Theta_4- \Theta_1=0$, 
 \[ \theta_{[1,4]}^+ + \theta_{[1,4]}^-= \ln \frac{k_3}{k_4-k_3}-2 \ln b + \ln \frac{k_2}{k_4-k_2}=0. \]
Then   
\beaa \tau_P &=&  \sqrt{k_2} e^{\theta_1+ \theta_2}+ \sqrt{\Delta_P} \sqrt{k_3} e^{\theta_1+ \theta_3}+ \sqrt{\Delta_P} \sqrt{k_2}e^{\theta_2+ \theta_4}\\
 &+&   \sqrt{k_3} e^{\theta_3+ \theta_4}. \eeaa
And $   \tau_P^{(1)} =   k_2 k_4  \sqrt{\Delta_P} \sqrt{k_2}e^{\theta_2+ \theta_4} +  k_3 k_4 \sqrt{k_3} e^{\theta_3+ \theta_4}. $ Then also from (\ref{amp})
\be  u=k_2- {\frac{(k_3-k_2)^2}{(\sqrt{k_4-k_3}+ \sqrt{k_4-k_2} )^2}}.   \label{kii} \ee
The last term is the interaction term between line soliton and kink soliton. It is different from the O-type soliton. In (\ref{ki}), the amplitude of interaction is higher than $k_2$; however, in (\ref{kii}), the amplitude of interaction is lower than $k_2$. Please also see the figure 3 .

\end{itemize}  
\section{Concluding Remarks}
In this article, we construct the non-singular soliton solutions of MKP-(II) using the Wronskian structure of $\tau$-functions. As a result, the totally non-negative Grassmannian manifold can be utilized to study the resonance of line solitons, as the KP-(II) solitons does.  Letting $k_1=0$, one can investigate the resonance of kink solitons. Also, Y-type kink-soliton resonance, O-type kink soliton and  P-type kink soliton of X-shape are investigated. The amplitudes of the intersections of  O-type and P-type are computed  after choosing appropriate phases and their lower bounds and upper bounds are estimated, and the ones of interactions of kink solitons and line solitons are also found. \\
\indent In addition, one makes a comparison with the KP-(II) equation. In MKP-(II) equation, all the parameters $k_i \geq 0$ to obtain non-singular soliton solutions; moreover, one can get multi-kink solitons when $k_1=0$.  Neither such condition nor any kink soliton exists for the KP-(II) equation. One has the resonance structure of the multi-kink solitons (\ref{pr}) and  (\ref{ree}). Therefore, in MKP-(II) equation one has different unbounded line solitons as $ y \to \pm \infty$ from the KP-(II) equation. As for the  O-type and P-type solitons, the interaction between kink solitons and line solitons could be interesting. The Mach-type soliton for the MKP-(II) could be interesting when compared with the KP-(II) equation \cite{ko5} and the Novikov-Veselov equation \cite{jh1}. Also, when $k_1$=0, the self-dual $\tau$-functions \cite{ko2} (or T-type soliton \cite{lp}) , which are characterized by identical sets of asymptotic line solitons as $ y \to \pm \infty$, are to be investigated. In particular,  the asymptotic line solitons inside the multi-kink soliton could be interesting. These issues will be published elsewhere. 

\subsection*{Acknowledgments}
This work is supported  by the Ministry of Science and Technology of Taiwan under Grant No. MOST 105-2115-M-606-001.

\end{document}